\documentstyle[preprint,aps,prl]{revtex}

\begin{document}
\preprint{SUSX-TH/96-002, hep-th/9602009}
\title{Vector-Tensor Duality}
\author{Amitabha Lahiri\thanks{lahiri@bose.ernet.in}}
\address{School of Mathematical \& Physical Sciences, \\ 
University of Sussex, Brighton BN1 9QH. U. K.\thanks{Current address:
S. N. Bose National Centre for Basic Sciences, Block JD, Sector III, 
Salt Lake City, Calcutta 700 091, W. B., INDIA.}}

\maketitle


\begin{abstract}
A dynamical non-abelian two-form potential gives masses to vector bosons 
via a topological coupling \cite{gvm}. Unlike in the abelian case, the 
two-form cannot be dualized to Goldstone bosons. Duality is restored by 
coupling a flat connection to the theory in a particular way, and the 
new action is then dualized to a non-linear sigma model. The presence of 
the flat connection is crucial, which saves the original mechanism of 
Higgs-free topological mass generation from being dualized to a sigma 
model.

\end{abstract}

The properties of an abelian two-form are well-known.  By itself, it
describes a massless particle \cite{opd}, while when coupled to an
abelian gauge field via a topological term it provides a gauge invariant
mass to the vector \cite{aurtak,trg,abl,minwar} without a residual
scalar Higgs.  The abelian mechanism was generalized to a compact
non-abelian gauge group sometime ago in the context of non-abelian
quantum hair on black holes \cite{qhair} and then as a way of giving
masses to vector bosons \cite{gvm}.  The non-abelian model acquires
significance when we consider the fact that the Higgs particle has not
been observed yet.  If it remains elusive in the coming generation of
accelerators, new ways of generating vector boson masses will have to be
considered.  The non-abelian two-form provides a plausible model for
Higgs-free mass generation.

The action for the dynamical non-abelian two-form is 
\begin{equation}
S = \int d^4x{\rm Tr}\bigg(-{1\over
12}H_{\mu\nu\lambda}H^{\mu\nu\lambda} - 
{1\over 4}F_{\mu\nu}F^{\mu\nu} +
{m\over 4}\epsilon^{\mu\nu\rho\lambda}B_{\mu\nu}F_{\rho\lambda}\bigg). 
\label{firstac}
\end{equation}
Here $D_\mu$ is the connection of some gauge group $SU(N)$, 
$H_{\mu\nu\lambda} = 
D_{[\mu}B_{\nu\lambda]} - \big[F_{[\mu\nu},
C_{\lambda]}\big]$ is the compensated field strength of a non-abelian 
two-form $B$
which lives in the adjoint representation,
$C$ is an auxiliary field which also lives in the adjoint 
representation, and $F_{\mu\nu} = [D_\mu, D_\nu]$ is the
field strength of the $SU(N)$ gauge field $A$. The presence of the 
compensating field $C$ is 
required \cite{gvm,thibau,hwalee,neto} in order to generalize the 
Kalb-Ramond 
transformation \cite{kr} to a non-abelian one, 
\begin{equation}
B_{\mu\nu} \to B_{\mu\nu} + D_{[\mu}\Lambda_{\nu]}; \qquad
C_\mu \to C_\mu + \Lambda_\mu.
\label{naKR}
\end{equation}
This symmetry of the action ensures that classically there are only
three degrees of freedom for each gauge index.  These can be thought
of as the three degrees of a massive gauge field, and there is no
degree of freedom corresponding to a residual Higgs field.  This model
can be quantized, and a pole appears in the propagator of the gauge
field when tree level diagrams are summed \cite{gvm,hwalee}, and there
is no residual Higgs particle.  A BRST-invariant quantum action for
this model has also been found recently \cite{hwalee,prep}, and it
seems possible that this model can be shown to be renormalizable and
unitary.

Any alternative for the Higgs mechanism must be shown to be unitary and
renormalizable in order for it to be given serious consideration.  There
are good reasons to think that the model is renormalizable -- it is by
power counting, and the propagators fall off as $1/k^2$.  On the other
hand, unitarity is less tractable in this model.  A calculation of
unitarity in scattering of longitudinal vectors from longitudinal
vectors runs into problems immediately because the longitudinal mode in
(\ref{firstac}) is not easily identifiable.  In the abelian model one 
can
write a duality relation of the form $H_{\mu\nu\lambda} =
\epsilon_{\mu\nu\lambda\rho}(mA^\rho + \partial^\rho\phi)$ where $\phi$
is a scalar.  This allows one to rewrite the action in an obviously
unitary form and it can be shown that unphysical modes do not propagate.
But no such duality relation exists for the non-abelian model 
(\ref{firstac}).  In this letter, I explore a modified version of this
model.  I shall show that when a flat connection couples to the model in
a particular way in addition to the dynamical gauge field already
present, the non-abelian antisymmetric tensor can be dualized to a
non-abelian version of the abelian duality relation.

To begin with, let me include an $SU(N)$ flat connection ${\tilde A}$ 
in the model in 
addition to the $SU(N)$ gauge field $A$, and define two vector 
fields 
${\cal A}_\mu$ and $\Phi_\mu$,
\begin{equation}
{\cal A}_\mu = {\displaystyle{1\over 2}}(A_\mu + {\tilde A}_\mu), \qquad 
\Phi_\mu = 
{\displaystyle{1\over 2}}(A_\mu - {\tilde A}_\mu).
\label{fielddefs}
\end{equation}
$\Phi_\mu$ transforms homogeneously under $SU(N)$ gauge transformations,
while ${\cal A}_\mu$ transforms like a connection.  
This allows the following
covariant derivatives and field strengths to be defined,
\begin{equation}
D_\mu = \partial_\mu + A_\mu;\qquad {\cal D}_\mu = \partial_\mu + 
{\cal A}_\mu;\qquad F_{\mu\nu} = [D_\mu, D_\nu];\qquad{\rm and}\ \ 
{\widetilde H}_{\mu\nu\lambda} = {\cal D}_{[\mu}B_{\nu\lambda]}.
\label{fstrdef}
\end{equation}
For the sake of simplicity I have ignored the auxiliary field (in other 
words set it to zero using the vector gauge transformation) in the 
definition of ${\widetilde H}$, but it can be restored without any 
problem. Now I 
can write down a modified version of the action (\ref{firstac}),
\begin{equation}
S =  \int d^4x{\rm Tr}\bigg(-{1\over
12}{\widetilde H}_{\mu\nu\lambda}{\widetilde H}^{\mu\nu\lambda} - 
{1\over 
4}F_{\mu\nu}F^{\mu\nu} +
{m\over 4}\epsilon^{\mu\nu\rho\lambda}B_{\mu\nu}F_{\rho\lambda}\bigg). 
\label{modac}
\end{equation} 

The vector-tensor duality shows up in the dynamics of this action. The 
equations of motion of the two dynamical fields $A_\mu$ and 
$B_{\mu\nu}$ as derived from this action are
\begin{eqnarray}
D_\nu F^{\nu\mu} - {1\over 4} [B_{\nu\lambda}, 
{\widetilde H}^{\mu\nu\lambda}] & + & {m\over 6} 
\epsilon^{\mu\nu\rho\lambda} 
D_{[\nu}B_{\rho\lambda]} = 0 \; , \\
{\cal D}_\lambda{\widetilde H}^{\mu\nu\lambda} + {m\over 2}
\epsilon^{\mu\nu\rho\lambda} F_{\rho\lambda} & = & 0 \; .
\label{eomB}
\end{eqnarray}
The second equation of this set is reduced to an identity by the ansatz
\begin{equation}
{\widetilde H}^{\mu\nu\lambda} = 
- 2m \epsilon^{\mu\nu\lambda\rho}\Phi_\rho.
\label{hansatz}
\end{equation}
Since ${\tilde A}$ is a flat connection, $[\partial_\lambda + {\tilde 
A}_\lambda, 
\partial_\rho + {\tilde A}_\rho] = 0$, it follows that 
\begin{eqnarray}
{\cal D}_\lambda{\widetilde H}^{\mu\nu\lambda} &=& - m 
\epsilon^{\mu\nu\lambda\rho} ({\cal D}_\lambda\Phi_\rho - 
{\cal D}_\rho\Phi_\lambda) \nonumber \\ &=& - {m\over 2} 
\epsilon^{\mu\nu\lambda\rho} 
\left(F_{\lambda\rho} - [\partial_\lambda + {\tilde A}_\lambda, 
\partial_\rho + 
{\tilde A}_\rho]\right)\nonumber \\ &=& - {m\over 2} 
\epsilon^{\mu\nu\lambda\rho} 
F_{\lambda\rho} \; .
\label{hsol}
\end{eqnarray}
So this ansatz solves the equation of motion (\ref{eomB}) for
$B_{\mu\nu}$. Although the fields appearing in this ansatz were already
present in the theory, this is not a constraint.  In fact, this ansatz solves
the Gauss' Law type constraint ${\cal D}_j \Pi_{ij} + {m\over
2}\epsilon_{ijk}F_{jk} \approx 0$. So one can quantize the theory after
eliminating this constraint by using the ansatz\footnote{For a given 
connection $A_\mu$, {\em any} flat connection $\tilde A_\mu$
will satisfy $(D + \tilde D)_{[\mu}(A - \tilde A)_{\nu]} = {1\over 
2}F_{\mu\nu}$. It follows that if $\Pi_{ij} = \epsilon_{ijk}(- 2m\Phi_k + 
K_k)$ is the general solution of the constraint equation, $K_k$ as 
a function of $\tilde A$ must satisfy ${\cal D}_{[j}K_{k]} = 0$ 
identically for all flat connections $\tilde A$. It can be shown 
that such a $K_i$ must vanish.}. However, this paper will be
limited to a classical analysis. This ansatz also simplifies the other
equation of motion considerably, as can be seen by rewriting the equation,
\begin{eqnarray}
D_\nu F^{\nu\mu} - {1\over 4}[B_{\nu\lambda}, 
{\widetilde H}^{\mu\nu\lambda}] + {m\over 6} 
\epsilon^{\mu\nu\rho\lambda} 
{\widetilde H}_{\nu\rho\lambda} + {m\over 6} 
\epsilon^{\mu\nu\rho\lambda} 
\left[\Phi_{[\nu}, B_{\rho\lambda]}\right] &=& \nonumber \\
D_\nu F^{\nu\mu} + {m\over 2} \epsilon^{\mu\nu\rho\lambda} 
[B_{\nu\lambda}, \Phi_\rho] - {m^2\over 3} \epsilon^{\mu\nu\rho\lambda} 
\epsilon_{\nu\rho\lambda\tau} \Phi^\tau + {m\over 2} 
\epsilon^{\mu\nu\rho\lambda} [\Phi_\nu, B_{\rho\lambda}] &=& \nonumber 
\\
D_\nu F^{\nu\mu} - 2m^2\Phi^\mu = D_\nu F^{\nu\mu} - m^2A^\mu + 
m^2{\tilde A}^\mu &=& 0,
\label{amotion}
\end{eqnarray}
which shows clearly that, at least at the classical level, the action
(\ref{modac}) describes a massive vector field. This gauge-invariant 
mass of the vector field can be obtained by rewriting the 
action  by substituting the duality ansatz  (\ref{hansatz}) 
into it. It is possible to rewrite the $B\wedge 
F$ interaction term using the flatness of ${\tilde A}$, 
\begin{eqnarray}
{m\over 4}\int{\rm Tr}\left( \epsilon^{\mu\nu\rho\lambda} 
B_{\mu\nu}F_{\rho\lambda}\right) &=& m\int{\rm Tr}\left( 
\epsilon^{\mu\nu\rho\lambda}B_{\mu\nu}{\cal D}_\rho\Phi_\lambda\right) 
\nonumber \\ &=& - m\int{\rm Tr}\left(\epsilon^{\mu\nu\rho\lambda}({\cal 
D}_\rho 
B_{\mu\nu})\Phi_\lambda\right) \nonumber \\ &=& -{m\over 3} 
\int{\rm Tr}\left( 
\epsilon^{\mu\nu\rho\lambda}{\widetilde 
H}_{\mu\nu\rho}\Phi_\lambda\right) 
\nonumber \\ &=& -4m^2\int{\rm Tr}\left(\Phi_\mu\Phi^\mu\right)
\label{bfah}
\end{eqnarray}
where I have neglected surface terms in the second line\footnote{The
surface term is of the form $\int dS^\mu
\epsilon^{\mu\nu\rho\lambda}B_{\nu\rho}\Phi_\lambda$.  This may have a
non-zero contribution in the presence of strings or monopoles if $B$
also carries a topological charge. In such a situation one also has to 
be
careful about substituting the ansatz into the action.}.  It should also
 be noted that
${\widetilde H}_{\mu\nu\lambda}{\widetilde H}^{\mu\nu\lambda} = - 24
m^2\Phi_\mu\Phi^\mu$.  When all this is put together, the action
(\ref{modac}) can be written as
\begin{equation}
S = \int d^4x{\rm Tr}\left(- {1\over 4}F_{\mu\nu}F^{\mu\nu} - 
2m^2 \Phi_\mu\Phi^\mu\right).
\label{phiac}
\end{equation}
As can be seen, all references to $B_{\mu\nu}$ has dropped out of the 
action. This action reproduces the equation of motion (\ref{amotion}) 
for a massive $A_\mu$.

Let me make a couple of digressions here.  As mentioned before, the
duality exists with minor modifications when the compensating
auxiliary field $C$ is introduced as in \cite{gvm}.  One has to
replace the field strength ${\widetilde H}$ by a compensated field
strength ${\widetilde H}'$,
\begin{equation}
{\widetilde H}'_{\mu\nu\lambda} = {\widetilde H}_{\mu\nu\lambda} - 
\left[F_{[\mu\nu}, C_{\lambda]}\right] + \left[\Phi_{[\mu}, 
D_{\nu}C_{\lambda]}\right],
\label{newh}
\end{equation}
and replace ${\widetilde H}$ by ${\widetilde H}'$ in the action, while
leaving the other terms as they were.  Then the non-abelian vector
gauge symmetry (\ref{naKR}) is a symmetry of the action.  It is easy
to see that the rest of the analysis above holds with the replacement
of ${\widetilde H}$ by ${\widetilde H}'$. Therefore, what has been done
in this paper boils down to a proof of classical equivalence of a 
St\" uckelberg-type theory with a particular theory of dynamical 
non-abelian two-forms. 

A clarification is needed on the issue of enforcing the flatness of
$\tilde A$ in the path integral. The action of Eq. (5) as written uses
$\tilde F_{\mu\nu} \equiv [\partial_\mu + \tilde A_\mu, \partial_\nu +
\tilde A_\nu] = 0$, which shows up in Eq. (9). In the path integral
this will be implemented by $\delta(\tilde F_{\mu\nu})$, which can be
rewritten as a term ${\rm Tr}({1\over 4}\epsilon^{\mu\nu\rho\lambda}
E_{\mu\nu}\tilde F_{\rho\lambda})$ in the action, where $E_{\mu\nu}$
is a $new$ Lie-algebra valued two-form. Let me also include the term
${\rm Tr}(-{m\over 4}\epsilon^{\mu\nu\rho\lambda}B_{\mu\nu}\tilde 
F_{\rho\lambda})$ which corresponds to what was set to zero in 
Eq.(\ref{hsol}). Then the total action is 
\begin{equation}
 S = \int d^4x {\rm Tr} \left[ -{1\over 12}\widetilde H_{\mu\nu\lambda}
\widetilde H^{\mu\nu\lambda} - {1\over 4}F_{\mu\nu}F^{\mu\nu}
 + {m\over 4}
\epsilon^{\mu\nu\rho\lambda} B_{\mu\nu}(F_{\rho\lambda} - \tilde 
F_{\rho\lambda}) + {1\over 4}\epsilon^{\mu\nu\rho\lambda}
E_{\mu\nu}\tilde F_{\rho\lambda}\right].
\label{unflat}
\end{equation}
The equation of motion for $B_{\mu\nu}$ which follows from this action 
is 
\begin{equation}
{\cal D}_\lambda\widetilde H^{\mu\nu\lambda} + {m\over 2}
\epsilon^{\mu\nu\rho\lambda} (F_{\rho\lambda} - \tilde F_{\rho\lambda})
= 0,
\label{neweom}
\end{equation}
and this is again solved by the ansatz $\widetilde H^{\mu\nu\lambda} =
- 2m \epsilon^{\mu\nu\lambda\rho}\Phi_\rho$, this time without the 
requirement of flatness $\tilde F_{\rho\lambda} = 0$. The analysis
proceeds as before, and we get the reduced action
\begin{equation}
S_{red} = \int d^4x {\rm Tr} \left[-{1\over 4} F_{\mu\nu}F^{\mu\nu} -
 2m^2\Phi_\mu\Phi^\mu + 
{1\over 4}\epsilon^{\mu\nu\rho\lambda} E_{\mu\nu}\tilde 
F_{\rho\lambda}\right].
\label{redac}
\end{equation}

In terms of path integrals, the action (\ref{modac}) corresponds to 
\begin{equation}
\int [{\cal D}A][{\cal D}\tilde A][{\cal D} B]\delta(\tilde 
F_{\rho\lambda})\exp(i\int {\rm Tr} (-{1\over 12}
\widetilde H^2 - {1\over 4}F^2
+ {m\over 2}B\wedge F)),
\label{modacpi}
\end{equation}
while the reduced action corresponds to
\begin{equation}
\int [{\cal D}A][{\cal D}\tilde A]\delta(\tilde
F_{\rho\lambda})\exp(i\int Tr (- {1\over 4}F^2 - 2m^2\Phi^2)).
\label{redacpi}
\end{equation}
In both the path integrals, the $\delta(F_{\rho\lambda})$
can be replaced by a term 
${\rm Tr} {1\over 4}\epsilon^{\mu\nu\rho\lambda}
 E_{\mu\nu}\tilde F_{\rho\lambda}$ in the Lagrangian. 
In other words, the duality relation makes it possible to 
integrate out the $B$-field without getting involved in the
subtleties in enforcing the flatness of $\tilde A$ at the
quantum level.

There are several things to be noticed about this duality, in
particular about the action (\ref{phiac}).  If one were to formulate
this system for an abelian gauge group one would find that the duality
relation (\ref{hansatz}) was in fact the well known duality for the
abelian model
\cite{abl}.  In other words, the model (\ref{modac}) is the non-abelian
generalization of the mass generation mechanism if one were to start 
from
the duality relation rather than from the action itself. Just as in the 
abelian case, the duality between $\Phi$ and the non-abelian two-form 
forbids any further interaction terms of mass dimension four involving
$\Phi$. In particular a $(\Phi^\mu\Phi_\mu)^2$ or similar interaction 
cannot be added to the action without destroying the duality. Nor can 
one add any kinetic terms for $\Phi$ (that is, other than 
$({\cal D}_{[\mu}\Phi_{\nu]})^2$, which can be rewritten as 
$(F_{\mu\nu})^2$).

Classically speaking, since ${\tilde A}$ is a flat connection, it is 
possible to find a $g$  such that 
${\tilde A}_\mu = - \partial_\mu gg^{-1}$. Then the action (\ref{phiac}) 
can be  written as 
\begin{eqnarray}
S &=& \int d^4x{\rm Tr}\left(-{1\over 4}F_{\mu\nu}F^{\mu\nu} - {m^2\over 
2}(A_\mu + \partial_\mu gg^{-1})^2\right) \nonumber \\ 
&=& 
\int d^4x{\rm Tr}\left(-{1\over 4}F_{\mu\nu}F^{\mu\nu} - {m^2\over 
2}(D_\mu g g^{-1})^2\right).
\label{sigma}
\end{eqnarray}
This shows that, at least classically, the system described by 
(\ref{phiac}) is in fact a gauged non-linear sigma model, which is
known to be non-renormalizable. However it would be 
inappropriate to dismiss the action (\ref{phiac}) (and by inference 
(\ref{modac})) as non-renormalizable or non-unitary.  In the case of 
(\ref{modac}), the duality relation (\ref{hansatz}) is highly non-local, 
which means that the longitudinal modes of massive gauge vectors in this 
model are related to the two-form through non-local relations. In the 
path integral quantization of the theory this implies a non-trivial 
Jacobian when a change of variables is made from $B_{\mu\nu}$ to 
$\Phi_\mu$. This transformation modifies the large momentum behaviour of 
the propagator and would introduce non-renormalizability.  In  canonical 
quantization the non-locality of the duality relation means that the 
amplitude of scattering longitudinal vectors off longitudinal vectors 
cannot be directly calculated from (\ref{modac}). In the case of 
(\ref{firstac}) things are even more complicated as no duality exists, 
and longitudinal vectors cannot even be related to the anitsymmetric 
tensor in a closed form. All attempts to prove or disprove tree-level 
unitarity of this model must therefore fail, and one has to look for 
alternative approaches, using BRST invariance for example 
\cite{hwalee,prep}, to quantize the theory in a self consistent manner.

There is yet another loophole which may protect the dual 
model (\ref{modac}) from non-unitarity and non-renomalizability.
The relation $\tilde A_\mu = - \partial_\mu gg^{-1}$ is a non-local 
relation which 
contributes a non-trivial Jacobian when one makes a change of variables 
in 
the path integral\footnote{I am indebted to A. Niemi for discussing and 
clarifying this point.}. It is then by no means obvious whether the 
model of 
(\ref{phiac}) is still identical to the non-linear sigma model as a 
quantum theory or if it escapes the fate of the latter through this 
loophole, which another model \cite{niemi} has tried to exploit 
recently. In that model the field strength $F$ in the interaction 
$B\wedge F$ is the 
field strength of the flat connection $\tilde A$, and $\Phi_\mu$, rather 
than a dynamical $B_{\mu\nu}$, is treated as a fundamental variable. The 
action proposed 
there is similar, but not identical, to the dualized action 
(\ref{phiac}). 
The main difference lies in the presence of  $(\Phi_\mu)^4$ 
interactions, 
and a symmetric kinetic term of the form $(D_{(\mu}\Phi_{\nu)})^2$ (the 
antisymmetric part can be rewritten as $(F_{\mu\nu})^2$). These two 
terms 
make sure that one cannot recover something like (\ref{modac}) by a 
duality transformation.

\centerline{\bf Acknowledgement}

It is a pleasure to thank T. J. Allen and A. Niemi for critically 
reading the manuscript. My thanks to the anonymous referee for 
asking for clarifications on several points which might have been
unobvious in a previous version.


\nonfrenchspacing


\end{document}